\documentclass[prl,twocolumn,superscriptaddress]{revtex4}

\usepackage{bm}
\usepackage{graphics}
\usepackage{amsmath,epsfig}
\usepackage{amssymb}
\usepackage{comment}

\begin{document}


\title{Experimental quantum coding against photon loss error}

\author{Chao-Yang Lu $^\star$}
\affiliation{Hefei National Laboratory for Physical Sciences at
Microscale and Department of Modern Physics, University of Science
and Technology of China, Hefei, Anhui 230026, P. R. China}

\author{Wei-Bo Gao $^\star$}
\affiliation{Hefei National Laboratory for Physical Sciences at
Microscale and Department of Modern Physics, University of Science
and Technology of China, Hefei, Anhui 230026, P. R. China}

\author{Jin Zhang}
\affiliation{Hefei National Laboratory for Physical Sciences at
Microscale and Department of Modern Physics, University of Science
and Technology of China, Hefei, Anhui 230026, P. R. China}

\author{Xiao-Qi Zhou}
\affiliation{Hefei National Laboratory for Physical Sciences at
Microscale and Department of Modern Physics, University of Science
and Technology of China, Hefei, Anhui 230026, P. R. China}

\author{Tao Yang}
\affiliation{Hefei National Laboratory for Physical Sciences at
Microscale and Department of Modern Physics, University of Science
and Technology of China, Hefei, Anhui 230026, P. R. China}

\author{Jian-Wei Pan $^\S\,$}
\affiliation{Hefei National Laboratory for Physical Sciences at
Microscale and Department of Modern Physics, University of Science
and Technology of China, Hefei, Anhui 230026, P. R. China}
\affiliation{Physikalisches Institut, Universit\"{a}t Heidelberg,
Philosophenweg 12, 69120 Heidelberg, Germany}

\date{\today}

\begin{abstract}
A significant obstacle for practical quantum computation is the loss
of physical qubits in quantum computers, a decoherence mechanism
most notably in optical systems. Here we experimentally demonstrate,
both in the quantum circuit model and in the one-way quantum
computer model, the smallest non-trivial quantum codes to tackle
this problem. In the experiment, we encode single-qubit input states
into highly-entangled multiparticle codewords, and we test their
ability to protect encoded quantum information from detected
one-qubit loss error. Our results prove the in-principle feasibility
of overcoming the qubit loss error by quantum codes.
\end{abstract}


\pacs{Valid PACS appear here}
\maketitle Quantum computers are expected to harness the strange
properties of quantum mechanics such as superposition and
entanglement for enhanced ways of information processing. However,
it has proved extremely difficult to build such devices in practice.
Arguably the most formidable hurdle is the unavoidable decoherence
caused by the coupling of the quantum computers to the environment
which destroys the fragile quantum information rapidly. It is thus
of crucial importance to find ways to reduce the decoherence and
carry out coherent quantum operations in the presence of noise.

Recent experiments have made progresses toward this goal by
demonstrating quantum error correction
\cite{cory,knillexperiment,ionexperiment,pittman,brien},
decoherence-free subspace \cite{dfs_optics,dfs_ion,steinberg,harald}
and entanglement purification \cite{puri_pan,puri_ion}. These
experiments were designed to cope with one special kind of
decoherence, that is, when qubits become entangled with the
environment or undergo unknown rotations in the qubit space. Such
errors can be represented as linear combinations of the standard
errors: no error, bit-flip, phase-flip, or both.


There is, however, another significant source of error --- the loss
of qubits in quantum computers. The qubit, which is the basic
element of standard quantum computation (QC), is supposed to be an
isolated two-level system consisting of a pair of orthonormal
quantum states. However, the majority of proposed quantum hardware
are in fact multi-level systems, and the states of qubits are
defined in a two-level subspace, which may leak out of the desired
qubit space and into a larger Hilbert space
\cite{preskill,wu,simon}. This problem is common in practical QC
with various qubits candidates, such as Josephson junctions
\cite{jose}, neutral atoms in optical lattices \cite{optical
lattice}, and, most notoriously, single photons which can be lost
during processing or owing to inefficient photon sources and
detectors \cite{rmp,klm,photonpra,knilldetector,photonloss}. The
loss of physical qubits is detrimental to QC since the working of
quantum gates, algorithms and error correction codes (see e.g.
\cite{shor,steane1,gottesman}) all hinge on the percept that the
quantum system remains in the qubit space.

Here we demonstrate the smallest meaningful quantum codes to protect
quantum information from detected one-qubit loss. Our experiment
deals with qubit loss in both the quantum circuit model and one-way
quantum computer model \cite{one-way}. We encode single-qubit states
into loss-tolerant codes which are multiparticle entangled states.
The performances of the quantum codes are tested by determining the
fidelities of the recovered states compared to the ideal original
states. Our results verify that the qubit loss error could in
principle be overcome by quantum codes.

\textbf{Theoretical schemes}


We now briefly review the quantum codes designed to tackle the
problem of qubit loss. A special class of quantum erasure-error
correction (QEEC) code was proposed by Grassl \textit{et al.}, where
a four-qubit code is sufficient to correct a detected one-qubit loss
error \cite{erasure}. The QEEC code was utilized by Knill \textit{et
al.} to deal with the photon-loss problem for scalable photonic QC
\cite{klm}. In recent years, extensive efforts have been devoted to
devising loss-tolerant quantum computer architectures
\cite{alifer,knillfaulttolerant,ralph,varnava}. In particular, in
the quantum circuit model Ralph \textit{et al.} employed an
incremental parity encoding method to achieve efficient linear
optics QC, and showed a loss-tolerant optical memory was possible
with a loss probability below 0.18 \cite{ralph}. In the new approach
known as the one-way QC, Varnava \textit{et al.} exploited the
inherent correlations in cluster states and introduced a novel
scheme for fault-tolerantly coping with losses in the one-way QC
that can tolerate up to 50\% qubit loss \cite{varnava}.

\textbf{QEEC codes}

To show the principle of the QEEC codes \cite{erasure}, let us start
with a specific example: a four-qubit code which is able to protect
a logical qubit from loss of a physical qubit. Here a logical qubit
$|\psi\rangle_l=a_0|0\rangle_l+a_1|1\rangle_l$ is encoded in the
subspace with four physical qubits as
\begin{eqnarray}
  |0\rangle_l=(|0\rangle_1|0\rangle_2+|1\rangle_1|1\rangle_2)(|0\rangle_3|0\rangle_4+|1\rangle_3|1\rangle_4)& &
  \\\nonumber
  |1\rangle_l=(|0\rangle_1|0\rangle_2-|1\rangle_1|1\rangle_2)(|0\rangle_3|0\rangle_4-|1\rangle_3|1\rangle_4)& &
\end{eqnarray}
This code can also be viewed as a combination of parity and
redundant encoding, which is the basic module in Ralph's scheme of
loss-tolerant optical QC \cite{ralph}.

We can consider the effect of a qubit loss as an unintended
measurement from which we learn no information. The main feature of
the code (1) is that the detected loss of any one of the physical
qubits will not destroy the information of the logical qubit, but
merely yields a recoverable Pauli error. Suppose for example, qubit
1 is lost. We first measure the qubit 2 in computational
($|0/1\rangle$) basis. With its measurement result ($q_2=0$ or $1$),
we can obtain a pure quantum state
$|\psi'\rangle_l=a_0(|0\rangle_3|0\rangle_4+|1\rangle_3|1\rangle_4)+(-1)^{q_2}a_1(|0\rangle_3|0\rangle_4-|1\rangle_3|1\rangle_4)$.
With similar reasoning, more-qubit loss can also be corrected by
increasing the size of loss-tolerant codes in the form of
$|\Psi\rangle_l=a_0(|0\rangle^{\otimes n}+|1\rangle^{\otimes
n})^{\otimes m}+b_0(|0\rangle^{\otimes n}-|1\rangle^{\otimes
n})^{\otimes m}$, which can be created \textit{e.g.}, by the
incremental encoding scheme proposed in Ref. \cite{ralph}.

\begin{figure}[t]
  \includegraphics[width=0.36\textwidth]{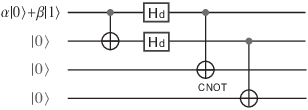}\\
  \caption{A quantum circuit with two Hadamard ($H_d$) gates and three CNOT gates for implementation of the four-qubit QEEC code.
  The stabilizer generators of the QEEC code are $X\otimes X\otimes X\otimes
  X$ and $Z\otimes Z\otimes Z\otimes Z$, where $X$ ($Z$) is short
  for Pauli matrix $\sigma_x$ ($\sigma_z$) \cite{gottesman}. As proposed by Vaidman \textit{et al.}, this four-qubit code
  can also be used for error detection \cite{vaidman}.
     }\label{}
\end{figure}
\textbf{Demonstration of the QEEC code}

A quantum circuit to implement the encoding of the four-qubit QEEC
code is shown in Fig. 1. To implement this, we design a linear
optics network (see Fig. 2A). The physical qubits are encoded by the
polarizations of photons, with $0$ corresponding to the horizontal
(H) polarization and $1$ to the vertical (V). As shown in ref.
\cite{dowling1,ralph2}, such an encoding method naturally
incorporates a loss detection mechanism and may enable high-fidelity
linear optical QC. Our experimental setup is illustrated in Fig. 2B.
We use spontaneous parametric down conversion \cite{kwiat} to create
the primary photonic qubits, which are then coherently manipulated
by linear optical elements to implement the coding circuit and read
out using single-photon detectors (see the caption of Fig. 2B and
Methods).

\begin{figure}[hbt]
\centering
  \includegraphics[width=0.5\textwidth]{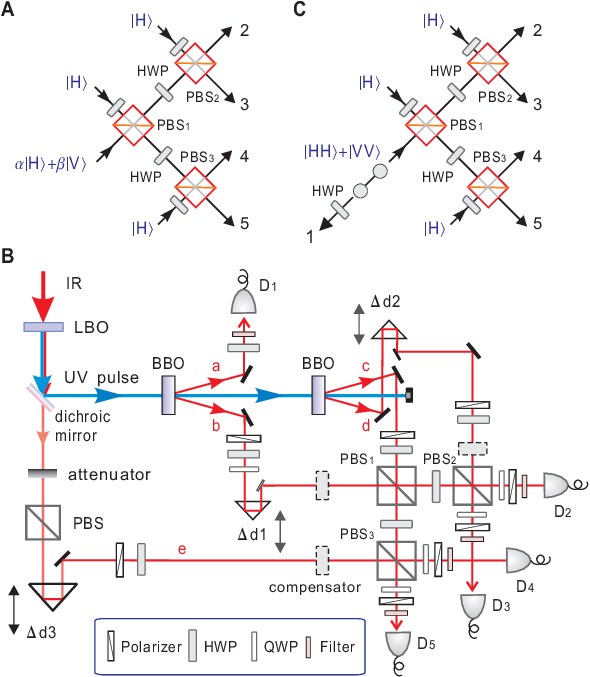}\\
  \caption{The linear optical networks and experimental setup.
  \textbf{A}. We simulate the CNOT gate in Fig. 1 using
a polarizing beam splitter (PBS) and a half-wave plate (HWP),
through which a control photon $(\alpha|H\rangle+\beta|V\rangle)$
and a target photon $|H\rangle$ evolve into
$(\alpha|H\rangle|H\rangle+\beta|V\rangle|V\rangle)$ after
postselection. Thus the circuit in Fig. 1 can be realized by this
linear optical network. \textbf{B}. A pulsed infrared laser (788nm,
120fs, 76MHz) passes through a LiB$_3$O$_5$ (LBO) crystal where the
laser is partially up-converted to ultraviolet ($\lambda$=394nm).
Behind the LBO, five dichroic mirrors (only one shown) are used to
separate the mixed ultraviolet (UV) and infrared light components.
The reflected UV laser passes through two $\beta$-barium borate
(BBO) crystals to produce two pairs of entangled photons. The
transmitted infrared laser is further attenuated to a weak coherent
photon source. To achieve good spatial and temporal overlap, the
photons are spectrally filtered by narrow-band filters
($\Delta\lambda_{\mathrm{FWHW}}=3.2\mathrm{nm}$, with peak
transmission rates of $\sim98\%$) and detected by fiber-coupled
single-photon detectors ($D_1$, $\cdots$, $D_5$) \cite{dik}. The
compensator consists of a HWP sandwiched by two thin BBO crystals.
By tilting the BBO, we can compensate the undesired phase shift in
the PBS. \textbf{C}. The five-photon cluster state can be prepared
by small modifications of the scheme of Fig. 2A.
     }\label{}
\end{figure}

\begin{figure*}[hbt]
\centering
  \includegraphics[width=0.75\textwidth]{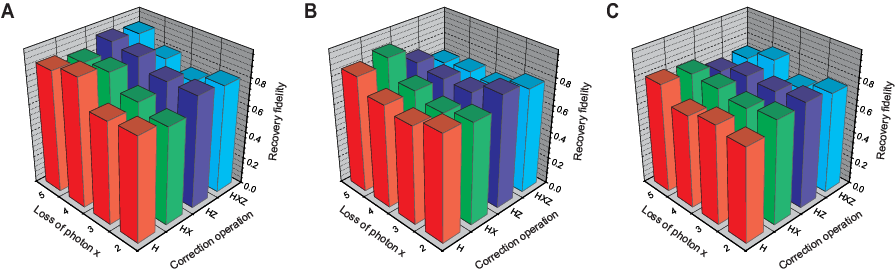}\\
  \caption{Experimental results of recovering quantum information
  from detected qubit loss. Recovered state
  fidelities are listed for all possible cases of photon loss ($2$ or $3$ or $4$ or $5$)
  and necessary feedforward correction operations ($H_d$ or $H_dX$ or $H_dZ$ or $H_dXZ$). \textbf{A}, data for input state
  $|V\rangle$.
  \textbf{B}, for $|+\rangle$. \textbf{C}, for $|R\rangle$.
     }\label{}
\end{figure*}

To demonstrate the quantum codes work for general unknown states, we
test three different input states: $|V\rangle$, $|+\rangle$, and
$|R\rangle=(|H\rangle+i|V\rangle)/\sqrt{2}$, which are encoded into
the four-qubit QEEC codes respectively as (normalizations omitted)
\begin{eqnarray}
 \nonumber |V\rangle_l &=&
(|H\rangle_2|H\rangle_3-|V\rangle_2|V\rangle_3)(|H\rangle_4|H\rangle_5-|V\rangle_4|V\rangle_5),
\\ \nonumber
  |+\rangle_l &=&
(|H\rangle_2|H\rangle_3|H\rangle_4|H\rangle_5+|V\rangle_2|V\rangle_3|V\rangle_4|V\rangle_5),\\
\nonumber
  |R\rangle_l &=&
(|H\rangle_2|H\rangle_3+|V\rangle_2|V\rangle_3)(|H\rangle_4|H\rangle_5+|V\rangle_4|V\rangle_5)\\
\nonumber &&
+i(|H\rangle_2|H\rangle_3-|V\rangle_2|V\rangle_3)(|H\rangle_4|H\rangle_5-|V\rangle_4|V\rangle_5),
\end{eqnarray}
where the subscript denotes the spatial mode. Interestingly they
show three distinct types of entanglement: $|V\rangle_l$ is a
product state of two Einstein-Podolsky-Rosen (EPR) pairs \cite{epr},
$|+\rangle_l$ is a four-qubit Greenberger-Horne-Zeilinger (GHZ)
state \cite{ghz}, while $|R\rangle_l$ is locally equivalent to a
cluster state \cite{cluster}.

We test the performance of the encoding process by determining
fidelities of the encoded four-qubit states. The fidelities are
judged by the overlap of the experimentally produced state with the
ideal one: $F=\langle\psi|\rho_{\mathrm{exp}}|\psi\rangle$. To do
so, we first decompose $\rho=|\psi\rangle\langle\psi|$ into locally
measurable observables which are products of Pauli operators [the
detailed constructions are shown in the supporting information (SI)
\textit{Methods}]. For the states $|V\rangle_l$, $|+\rangle_l$ and
$|R\rangle_l$ we need to take 9, 5 and 9 settings of four-photon
polarization correlation measurements respectively, each composed of
$2^4$ coincidence detections to determine the probabilities of
different outcome combinations. From the data shown in the SI
\textit{figures}, the fidelities of the QEEC codewords are:
$F_V=0.620\pm0.017$, $F_+=0.566\pm0.020$, $F_R=0.554\pm0.017$. The
fidelities of the four-qubit GHZ state and cluster state are above
the threshold of 0.5, thus they are confirmed to contain genuine
four-partite entanglement \cite{witness,toth}. The imperfections of
the fidelities are caused mainly by high-order emissions of
entangled photons and remaining distinguishability of independent
photons overlapping on the PBSs. Finally, it should be noted that,
as proposed in Ref. \cite{knillexperiment}, for the purpose of
``benchmarking'' quantum computers, more settings of measurements
will be needed to infer the average
fidelity of the quantum coding. 

\textbf{``Loss-and-Recovery'' test}

Now we test the codes' ability to protect the logical qubit
information from one detected physical qubit loss through a
``loss-and-recovery'' process. Here we simulate the loss of a photon
by detecting the photon without knowing its polarization
information, which only tells us that the photon is lost.
Experimentally, this is done by placing no polarizer or PBS in front
of the detector.

In principle, the QEEC code works when only one and any one of the
four physical qubits is lost. In our experiment, we test
individually all possible cases where any single one of the four
photons is lost. For instance, if we assume photon 2 is lost, the
experimental procedure goes as follows. We erase the photon 2,
perform a measurement in $H/V$ basis on photon 3 and in $+/-$ basis
on photon 4. Depending on different measurement results:
$|H\rangle_3|+\rangle_4$, $|V\rangle_3|+\rangle_4$,
$|H\rangle_3|-\rangle_4$, $|V\rangle_3|-\rangle_4$, correction
operations: $H_d$, $H_dX$, $H_dZ$, $H_dXZ$ should be applied on
photon 5. As a proof-of-principle, here we apply corrections for
every individual outcome of photon 3 and 4, and determine the state
fidelity of photon 5 compared to the original input state. For an
explicit example, if we fix the polarizers in front of D$_3$ and
D$_4$ in $|V\rangle$ and $|+\rangle$ polarization, we should apply
$H_d X$ to photon 5 and then measure its fidelity. Each measurement
is flagged by a fivefold coincidence event where all five detectors
fire simultaneously.

Figure 3 shows the measured recovery fidelities for all possible
combinations. For input states $|V\rangle$, $|+\rangle$ and
$|R\rangle$, the recovery fidelities averaged over all possible
measurement outcomes are found to be $0.832\pm0.012$,
$0.764\pm0.014$, and $0.745\pm0.015$ respectively, which well prove
the effectiveness of the four-qubit QEEC codes. It can be noticed
that the encoding and recovery fidelities for the state $|V\rangle$
is higher than those for $|+\rangle$ and $|R\rangle$. We believe
this is because in our setup, the coding process for $|V\rangle$
requires interference of photons only on PBS$_2$ and PBS$_3$ whilst
for the latter cases the interferences involve all three PBSs. Also
it can be seen from Fig. 3A that for the input state $|V\rangle$,
the recovery fidelities when we simulate photon $4$ or $5$ is lost
are considerably better than those when photon $2$ or $3$ is lost.
We note this is because in the former case, the interference
involves with dependent photons $c$ and $d$ (from the same EPR pair)
on the PBS$_2$ whilst the latter case requires interference of
independent photons $b$ and $e$ on the PBS$_3$.

\textbf{Loss-tolerant one-way QC}

Now we consider how to overcome the qubit loss in the one-way QC
model \cite{one-way}. In this model, QC is achieved by single-qubit
measurements on prepared highly-entangled cluster states
\cite{cluster}, where the orders and choices of measurements
determine the algorithm computed. It is important to note that the
loss-detection is naturally incorporated in the measurement step in
this QC model.

To tackle fault tolerance in this architecture, novel protocols have
been developed by exploiting the built-in properties of the
entangled cluster states which provide natural resilience to
decoherence \cite{raussendorf,varnava}. In particular, Varnava
\textit{et al.} utilized the tree-shaped graph states and analyzed
that a high error threshold of $0.5$ exists for qubit loss error
\cite{varnava}. The salient feature of this scheme is illustrated in
Fig. 4. Briefly, the tree graph state takes advantage of the perfect
correlations in the cluster states and embodies two useful features
that enable reduction of the effective qubit loss rate. First, it
allows multiple attempts to do the desired measurement on an encoded
qubit. Second, it is designed such that any given qubit within the
cluster could be removed (see the caption of Fig. 4).

\begin{figure}[t]
\centering
  \includegraphics[width=0.49\textwidth]{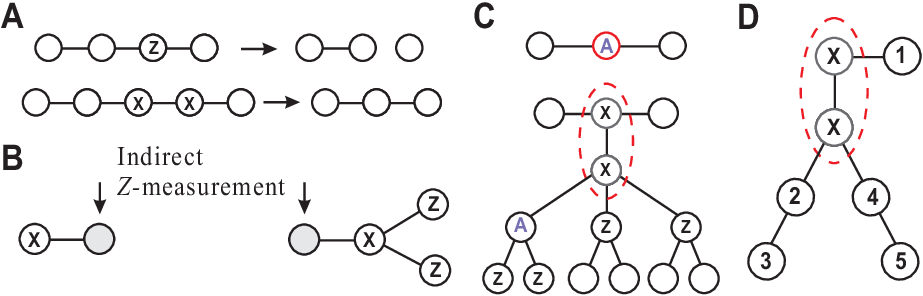}\\
  \caption{Principle of the loss-tolerant one-way quantum
  computation scheme \cite{varnava}. A cluster state can be represented by a graph,
  where the vertices take the role of qubits while the edges represent
  interaction \cite{graph}.
  \textbf{A}. Certain measurements on a cluster state have interesting effects:
  A $Z$ measurement removes the qubit from
the cluster and breaks all the bonds between the qubit and the rest;
Two adjacent $X$ measurements on a linear cluster remove the qubit
and form direct bonds between their neighbors. \textbf{B}. A cluster
state is an eigenstate of a set of stabilizers, which predict with
certainty correlations in the measurement outcomes of certain sets
of measurements. For instance, for a two-qubit state that is
stabilized by the operator $X_1Z_2$, if observable $X_1$ is measured
then the outcome of $Z_2$ is known with certainty. This allows us to
measure a qubit even if it is lost, which is called ``indirect
measurement" \cite{varnava}. \textbf{C}. The tree-graph cluster
state which can be used for reduction of the effective qubit loss
rate. We plant a cluster tree by two adjacent $X$ measurements, on
which instead of doing the $A$ measurement on the in-line qubit, we
can perform the measurement on a qubit in the third horizontal
level. When this measurement succeeds, we break the bounds with all
other qubits in the tree. If it fails, we remove this damaged qubit
and attempt the $A$ measurement on other qubits in the third level.
The tree structure ensures that the removal of damaged qubits can be
done by direct or indirect $Z$ measurements. \textbf{D}. The
five-qubit cluster state can be used for an in-principle
verification of this scheme.}\label{}
\end{figure}

\textbf{Creation of a five-qubit cluster state}

We use a five-qubit cluster state for demonstration of Varnava
\textit{et al.}'s scheme \cite{varnava}. As shown in Fig. 4D, it can
be thought of as being reduced from a seven-qubit cluster after two
adjacent $X$ measurements. Alternatively, the cluster can be grown
directly, as we do in our experiment.

Starting from an EPR pair and three single photons, we use the
linear optical network shown in Fig. 2C and the experimental setup
shown in Fig. 2B to create the five-photon cluster state (see
Methods). When each of the five output modes registers a photon, the
five photons are in the highly-entangled cluster state
\begin{eqnarray}\label{five cluster}
\nonumber|\phi_5\rangle=\frac{1}{2}\,(|H\rangle_1|H\rangle_2|H\rangle_3|H\rangle_4|H\rangle_5+|H\rangle_1|V\rangle_2|V\rangle_3|V\rangle_4|V\rangle_5
\\ \nonumber
\,+\,|V\rangle_1|H\rangle_2|H\rangle_3|V\rangle_4|V\rangle_5+|V\rangle_1|V\rangle_2|V\rangle_3|H\rangle_4|H\rangle_5),
\end{eqnarray}
where the subscript labels the photon's spatial mode (see Fig. 2C).
The state $|\phi_5\rangle$ is local unitarily equivalent to the
five-qubit linear cluster state shown in Fig. 4D under the $H_d$
transformations on photon $1$, $3$ and $5$. This is, to our best
knowledge, the longest one-dimensional cluster state realized so
far.

\begin{figure}[t]
\centering
  \includegraphics[width=0.48\textwidth]{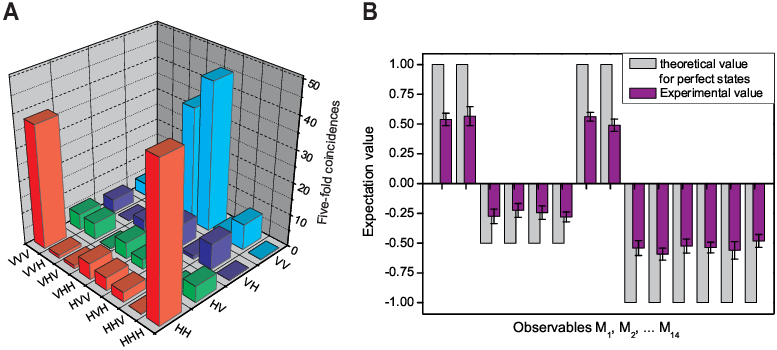}\\
  \caption{Experimental results of the five-photon cluster state $|\phi_5\rangle$.
  \textbf{A}. Five-photon detection events in the $H/V$ basis.
  \textbf{B}. Measured expectation values of the other 14 observables
  $M_1$, $M_2$, $\cdots$, $M_{14}$ (detailed representations shown in SI \textit{Methods}) to determine the fidelity of the cluster state $|\phi_5\rangle$. The error
  bars denote one standard deviation, deduced from propagated
  Poissonian counting statistics of the raw detection events.
     }\label{}
\end{figure}

To determine the fidelity of the five-photon cluster state, we
decompose the projector $|\phi_5\rangle\langle\phi_5|$ into $15$
local measurable observables (see the SI \textit{Methods}), each
takes $2^5$ five-fold coincidence measurements. The experimental
results are shown in Fig. 5, yielding
$F_{c}=\langle\phi_5|\rho_{exp}|\phi_5\rangle=0.564\pm0.015$. As the
fidelity of the cluster state exceeds $0.5$, the presence of true
five-partite entanglement of our cluster state is also confirmed
\cite{toth}.


\textbf{One-way QC in the presence of loss}

With the cluster state prepared, now we demonstrate its
loss-tolerant feature by simulation of a quantum circuit in the
presence of loss. First let us briefly review how QC is done by
measurements in the one-way model. The measurement is chosen in
basis $B_j(\alpha)=\{|+\alpha\rangle_j,|-\alpha\rangle_j\}$, where
$|\pm\alpha\rangle_j=(|0\rangle_j\pm
e^{i\alpha}|1\rangle_j)/\sqrt{2}$, which realizes the single-qubit
rotation $R_z(\alpha)=\mathrm{exp}(-i\alpha \sigma_z/2)$ followed by
a Hadamard operation on the encoded qubit in the cluster. We define
the outcome $s_j=0$ if the measurement on the physical qubit $j$
yields $|+\alpha\rangle_j$, and $s_j=1$ if it is
$|-\alpha\rangle_j$. When $s_j=0$, the computation proceeds without
error, whereas when $s_j=1$, a known Pauli error is introduced that
has to be compensated for (see ref. \cite{one-way,walther,robert}
for more details).

The two-qubit cluster shown in Fig. 6A can implement a simple
circuit, rotating an encoded input qubit $|+\rangle$ to an output
state: $|\psi_{\mathrm{out}}\rangle=X^{s_a}H_dR_z(\alpha)|+\rangle$.
With this two-qubit cluster, however, one can only have a one-shot
$A$ measurement on the qubit $a$, that is, if this measurement fails
then the whole computation fails. As a comparison, the five-qubit
cluster state we prepared can be used to realize the circuit in a
more robust fashion. It provides two alternative and equivalent
attempts to do the $A$ measurement as depicted in Fig. 6B and 6C.
And if any one of the qubits ($2$, $3$, $4$, $5$) for the $A$
measurement is lost, we can always find a suitable indirect $Z$
measurement to remove the damaged qubit. For example, if the $A$
measurement on qubit 2 fails, we can try to remove it from the
cluster by an indirect $Z$ measurement, and then proceed to make the
$A$ measurement on qubit 4. It can be checked that as long as no
more than one physical qubit is lost, the computation will be
successful.

\begin{figure}[t]
\centering
  \includegraphics[width=0.44\textwidth]{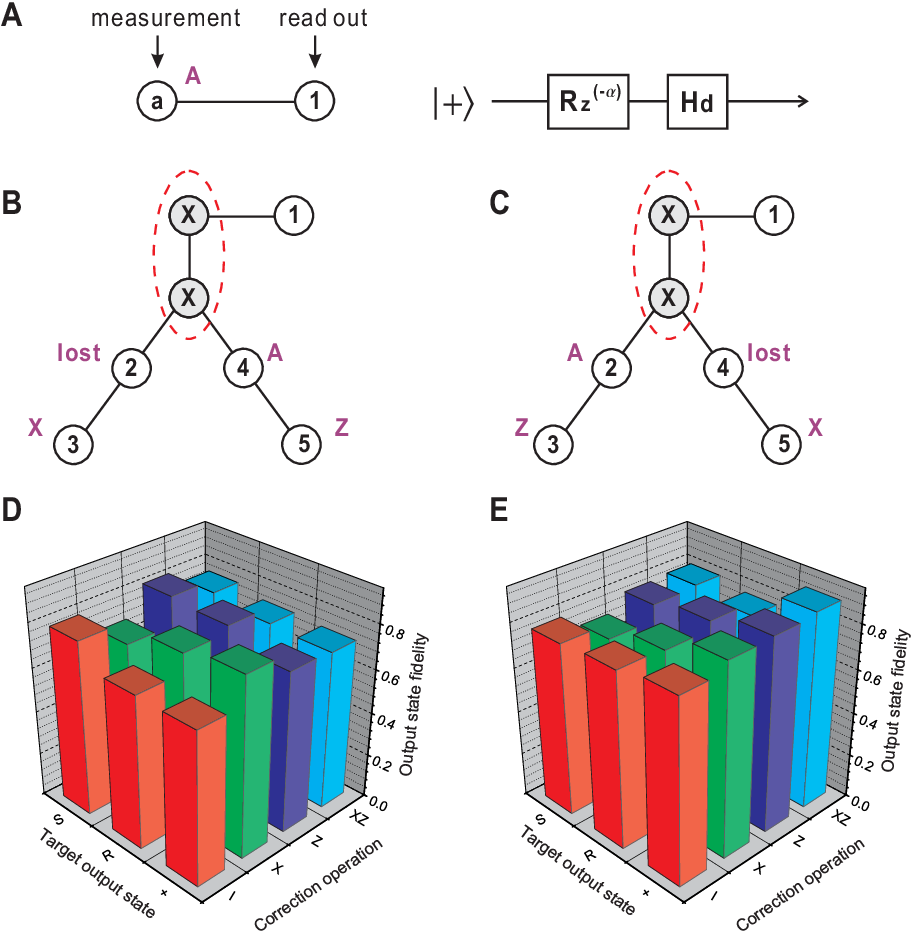}\\
  \caption{Experimental results of loss-tolerant one-way quantum
  computing. \textbf{A}. A two-qubit cluster state used to simulate a
  single-qubit rotation circuit by a measurement on qubit $a$.
  \textbf{B-C}. A five-qubit cluster state could realize the
  circuit in the presence of one-qubit
  loss. \textbf{D-E}. The experimentally measured fidelities of output
  states of the single-qubit rotation circuit. \textbf{D} (\textbf{E}) shows
  the results of the scheme \textbf{B} (\textbf{C}) respectively.
  Measurements on the qubit 4 (2) are performed in basis
  $B_j{(\alpha)}$ for different $\alpha$ value, \{$0, -\pi/2, -\pi/3$\} so that the target
  output state will be \{$|+\rangle$, $|R\rangle$, $|S\rangle$\}
  respectively.
     }\label{}
\end{figure}

Now we demonstrate it experimentally. To verify the scheme depicted
in Fig. 6B, we erase photon $2$, which makes the remaining cluster
in a mixed state. Then we make a $X$ measurement (which corresponds
to the $H/V$ basis in the laboratory basis for the actual state
$|\phi_5\rangle$) on photon $3$
--- this should effectively remove the loss error out of the cluster,
leaving it as a smaller but pure quantum cluster state. Next the
redundant photon $5$ is measured in the $Z$ basis (which corresponds
to the laboratory basis $+/-$). Depending on the measurement
outcomes of photon $3$ and $5$ ($|H\rangle_3|+\rangle_5$,
$|V\rangle_3|+\rangle_5$, $|H\rangle_3|-\rangle_5$,
$|V\rangle_3|-\rangle_5$), Pauli corrections ($I$, $X$, $Z$, $XZ$)
are applied to photon $4$. After that, measurements in the basis
$B_j(\alpha)$ are applied on photon $4$ to implement the rotation.
We choose $\alpha$ to be three different values, $0$, $-\pi/2$,
$-\pi/3$ so that theoretically the output states will be
$|+\rangle$, $|R\rangle$ and
$|S\rangle=(|0\rangle+e^{i\pi/3}|1\rangle)/\sqrt{2}$, respectively.
Then we readout the polarizations of photon $1$ and determine its
fidelities compared to the ideal states. The scheme of Fig. 6C is
also tested in a similar manner.

In Fig. 6D, 6E we show the experimental results of one-way QC in the
presence of one-qubit loss. In the case of photon $2$ lost, we find
an average fidelity of $0.738\pm0.029$, $0.750\pm0.030$, and
$0.765\pm0.028$ for target output state $|+\rangle$, $|R\rangle$ and
$|S\rangle$ respectively. In the case of photon $4$ lost, the
average fidelity is $0.865\pm0.021$, $0.792\pm0.029$, and
$0.767\pm0.030$ for target output state $|+\rangle$, $|R\rangle$ and
$|S\rangle$. Here the difference of the fidelity performance is
caused by similar reasons as in the QEEC codes. For instance, the
case for target output state $|+\rangle$ corresponds to Fig. 3A
where the input state is in the state of $|V\rangle$. These results
conclusively demonstrate
the underlying principle of loss-tolerant one-way QC. 

\textbf{Discussion}

As in all current linear optical QC experiments, the multiphoton
code states here are created probabilistically and conditioned on
that there is one and only one photon out of each output, a
technique called post-selection \cite{dik}. While this does not
prevent an in-principle verification of the loss-tolerant quantum
codes, we note eventual scalable implementations will need
significant improvements such as on-demand entangled photon sources
and high-efficiency single-photon detectors.




In summary, we have demonstrated both in the quantum circuit model
and in the one-way model, the smallest meaningful quantum codes to
protect quantum information from qubit loss error. These quantum
codes are the key modules for the loss-tolerant quantum computer
architectures \cite{ralph,varnava} and can in principle be extended
to larger number of qubits. Our results verify that it is possible
to overcome the qubit loss error, a major decoherence mechanism
common in many physical systems, and thus constitute a necessary
step toward scalable quantum information processing, especially for
photon-based QC. The loss-tolerant quantum codes can be further
concatenated with standard quantum error correction codes
\cite{shor,steane1,gottesman} or decoherence free space
\cite{dfs_optics,dfs_ion,steinberg,harald} to correct multiple
errors, and may become a useful part for future implementations of
quantum algorithms
\cite{walther,robert,lieven,kwiatsearch,white,luprl}.

\textbf{Methods}

\textbf{Experimental implementation} We first prepare two entangled
photon pairs in spatial modes $a$-$b$ and $c$-$d$ with an average
coincidence count of $\sim6.2\times10^4s^{-1}$, and a pseudo-single
photon source in mode $e$, which has a very small probability ($p$)
of containing a single photon for each pulse. The $p$ value is
carefully chosen to be $\sim0.07$ experimentally to get an optimal
visibility \cite{rarity}. Photon $a$ serves as a trigger, indicating
that the pair photon $b$ is under way; Photon $b$ is prepared in
different input states; Photon $c$, $d$, $e$ are initialized in the
$+45^{\circ}$ linear polarization state
$|+\rangle=(|H\rangle+|V\rangle)/\sqrt{2}$. We superpose photon $b$,
$c$, $d$, $e$ on the three PBSs step by step. For alignment on the
PBS$_1$, for instance, we first initialize the two photons $b$ and
$c$ at state $|+\rangle$ before they enter into the PBS$_1$, and, by
making fine adjustment of delay $\Delta d_1$, we are able to observe
the two-photon Hong-Ou-Mandel type dip \cite{hom} after the PBS$_1$
by performing polarization measurements in both output modes in the
$|\pm\rangle=(|H\rangle\pm|V\rangle)/\sqrt{2}$ basis. Similarly,
optimal superpositions of the photons on the other two PBSs are also
achieved. The optimal interference occurs at the regime of zero
delay, where our experimental measurements are performed. Note some
PBSs may have undesired phase shifts, that is, when two input
photons that prepared in the states of $|+\rangle$ are superposed on
the PBSs, the output photons are not in the expected state of
$(|H\rangle|H\rangle+|V\rangle|V\rangle)/\sqrt{2}$ but in the state
of $(|H\rangle|H\rangle+e^{i\theta}|V\rangle|V\rangle)/\sqrt{2}$
with a $\theta$ phase shift. This can be overcome by a compensator
(see Fig. 2B) which can introduce a phase delay between $|H\rangle$
and $|V\rangle$.

\textbf{Five-photon cluster state preparation} An EPR photon pair is
prepared in the spatial mode $a$-$b$ in the state
$(|H\rangle_a|H\rangle_b+|V\rangle_a|V\rangle_b)/\sqrt{2}$ with a
visibility of $92\%$ in the $+/-$ basis. We note that this
non-perfect visibility may add additional noise into the five-photon
cluster state compared to the four-photon states using the scheme as
shown in Fig. 2A. To get a better fidelity for the cluster state, we
lower the pump power and obtain an average coincidence of
$\sim5.0\times10^{4}s^{-1}$ in modes $a$-$b$ and $c$-$d$, such that
the rate of double pair emission of entangled photons is diminished.
Meanwhile, the $p$ value for the pseudo-single photon source is also
reduced to $\sim0.06$ accordingly.

\textit{Acknowledgements}: We thank T. Rudolph, D. Browne, O.
G\"{u}hne, T. Ralph and M. Zukowski for helpful discussions. This
work was supported by the NNSF of China, the CAS and the National
Fundamental Research Program (under Grant No. 2006CB921900). This
work was also supported by the Alexander von Humboldt Foundation and
Marie Curie Excellence Grant of the EU.

~\\
quantum computation $\mid$ quantum error correction $\mid$ ~~~~~~~~~
multi-photon entanglement

~\\
Author contributions: C.-Y.L., W.-B.G. and J.-W.P. designed
research, all authors performed the experiment, C.-Y.L. and J.-W.P.
wrote the paper.

This article is a PNAS Direct Submission.

The authors declare no conflict of interest.

 $^{\star}$ These authors
contributed equally to this work.

$^{\S}$ Correspondence and request for materials should be sent to
C.-Y.L (cylu@mail.ustc.edu.cn) and J.-W.P
(jian-wei.pan@physi.uni-heidelberg.de).


\end{document}